\documentclass[runningheads]{llncs}
\usepackage{graphicx}
\usepackage{subcaption}
\usepackage{todonotes}
\usepackage{siunitx}
\usepackage[nolist]{acronym}
\usepackage{comment}
\usepackage{siunitx}  
\usepackage{xcolor}
\usepackage{tcolorbox} 
\usepackage{url}
\begin{document}
\addtolength{\textfloatsep}{-4mm}
\addtolength{\floatsep}{-2mm} 

%
\title{Combining Static and Dynamic Traffic with Delay Guarantees in Time-Sensitive Networking}
\titlerunning{Combining Static and Dynamic Traffic with Delay Guarantees in TSN}
%
\author{Lisa Maile\orcidID{0000-0002-2528-7760} \and
Kai-Steffen Hielscher\and
Reinhard German}
\institute{Computer Networks and Communication Systems\\Friedrich-Alexander-Universität Erlangen-Nürnberg, Germany }
%
\maketitle              
\vspace{-2mm}
\begin{abstract}
To support reliable and low-latency communication, Time-Sensitive Networking introduced protocols and interfaces for resource allocation in Ethernet. However, the implementation of these allocation algorithms has not yet been covered by the standards. Our work focuses on deadline-guaranteeing resource allocation for networks with static and dynamic traffic. To achieve this, we combine offline network optimization heuristics with online admission control and, thus, allow for new flow registrations while the network is running. We demonstrate our solution on Credit-Based Shaper networks by using the delay analysis framework Network Calculus. We compare our approach with an intuitive and a brute-force algorithm, where we can achieve significant improvements, both, in terms of quality and runtime. Thereby, our results show that we can guarantee maximum end-to-end delays and also increase the flexibility of the network while requiring only minimal user input.

\end{abstract}

\begin{table}[b]
	\vspace{-4mm}
	\centering
	\begin{minipage}{0.95\linewidth}
		\vspace{1em}
		\begin{tcolorbox}[colback=blue!5!white, colframe=blue!40!black, boxrule=0.25pt]
				\scriptsize This version of the article has been accepted for publication, after peer review and is subject to Springer Nature’s AM terms of use, but is not the Version of Record and does not reflect post-acceptance improvements, or any corrections. The Version of Record is available online at: \url{https://doi.org/10.1007/978-3-031-48885-6_8}\end{tcolorbox}
	\end{minipage}
\end{table}
\section{Introduction}
With the increasing need for ultra-reliable and low latency communication, there has been a growing demand for more advanced transmission mediums to support them. As a response, the IEEE \ac{TSN} task group developed new standards that, i.a., allow for resource allocation in Ethernet with the introduction of central configuration units and decentralized reservation protocols. 
However, the current standards do not provide guidance on the configuration of \ac{TSN} networks. Therefore, several works have addressed this by proposing either offline (e.g.,~\cite{craciunas_scheduling_2016,gavrilut_traffic-type_2020}) or online (e.g.,~\cite{grigorjew_decentralSP_2020,maile_journal_2022}) configuration approaches. However, offline solutions do not allow for variable traffic and dynamic flows during the runtime of the network, while online solutions require the a-priori definition of non-trivial delay bounds by a user. To this end, we introduce a novel tool that combines offline and online configuration to support static and dynamic traffic while providing delay guarantees for time-sensitive flows. Thereby, offline configuration optimizes network setup in advance, while online configuration enables flow reservation and de-reservation while the network is running.

Combining both offline and online resource reservation with delay constraints is crucial in the design and operation of Industry 4.0 networking systems.
Firstly, offline optimization allows for the creation of an optimal plan for resource allocation and scheduling. This ensures that the network is designed to operate at peak efficiency and that resources are allocated effectively.
However, network conditions are rarely static and can change over time due to varying traffic patterns, new applications, devices, or other factors. Online admission control allows the network to respond quickly to changing conditions while maintaining its performance.

In this paper, we present a new approach that combines offline and online resource reservation for time-sensitive communication. The goal is to offer an auto-configuration framework which eliminates the need for manual intervention while still offering reliable flow delays.
To the best of our knowledge, we are the first to propose a combined solution for offline and online configuration for \ac{TSN} with safe delay guarantees. 

The remainder of this paper is structured as follows. Section~\ref{sec:problem} introduces related work and motivates the problem. Section~\ref{sec:TSN} presents an overview of our framework and its relation to the TSN standards. We explain our heuristic optimization in Section~\ref{sec:solution}.  Afterwards, we extensively evaluate our approach in Section~\ref{sec:evaluation}. Finally, Section~\ref{sec:conclusion} concludes this paper.

\section{Problem Definition and Related Work}
\label{sec:problem}
Traditional deadline-constrained optimization approaches take a set of flows as input and determine the optimal configuration of network resources, such as bandwidth or time-slots, as output. For example, when using time-triggered gates, individual time-slots are assigned for each flow, e.g.,~\cite{craciunas_scheduling_2016,anna}. For other schedulers, such as the \ac{CBS}, optimization approaches optimize the reserved bandwidth per traffic class, called idleSlope, so that all flows keep their deadline constraints, e.g., in~\cite{li_sdn-based_2019,gavrilut_traffic-type_2020}. 
However, all of these approaches have in common that, when additional time-sensitive flows are added to the network later, the reserved resources, such as time-slots or bandwidth, need to be adapted. This affects existing flow reservations, which evokes the necessity to re-validate all reservations, making these approaches inapplicable for dynamic networks.

To avoid this, flexible online solutions have been proposed in the last years, such as~\cite{Guck,maile_journal_2022,grigorjew_decentralSP_2020,maile_decentral_2023}.
They offer the opportunity to add and remove flow reservations while the network is running and still provide safe deadline guarantees. The underlying idea is that, instead of configuring resources such as time-slots or bandwidth, each hop is configured with priority-dependent maximum delay bounds. When the network changes, each hop is allowed to change its configuration, e.g., adapting the reserved bandwidth. Thereby, for each hop, it is validated that the new configuration does not violate the pre-configured delay bounds. This validation can either be done in a central controller instance, such as in~\cite{Guck,maile_journal_2022}, or with a decentralized reservation protocol, as proposed by~\cite{grigorjew_decentralSP_2020,maile_decentral_2023}.

\begin{figure}[!t]
		\centering\includegraphics[width=0.6\linewidth]{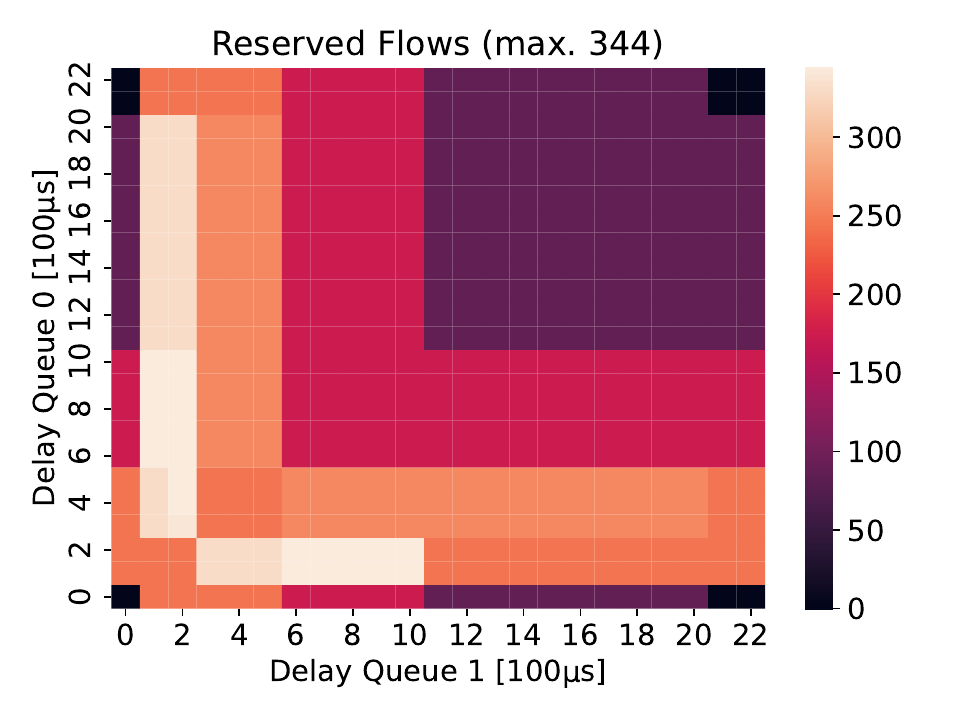}
		\caption{Impact of delay bounds for individual priority queues on the number of successful flow reservations.}\label{fig:heatmap}
\end{figure}

While online solutions offer high flexibility, we showed in~\cite{maile_journal_2022} that the choice of pre-configured delay bounds highly influences the possibility of reserving flows. We illustrate this in Fig.~\ref{fig:heatmap}, where we used the approach in~\cite{maile_journal_2022} to reserve flows. We assume a single-hop with two priority queues and a maximum of 344 flows\footnote{We defined the flows according to the traffic profiles 1, 2, 3, and 4 as defined in Table~\ref{table:profiles}, with 86 flows per profile and a bandwidth of \SI[per-mode=symbol]{1}{\giga\bit\per\second}.} for this example. Fig.~\ref{fig:heatmap} varies the delay bounds for the two priority queues. We can observe that their choice is of uttermost importance and non-trivial. High delay bounds would lead to a violation of the flows' deadlines, whereas low delay bounds require the reservation of more resources (e.g., more bandwidth to ensure a faster transmission) and, thus, allow for fewer flow reservations in total. Thus, the optimal delay bound values have to be chosen carefully and must neither be too strict nor too loose. Deriving the best delay bounds gets even more complex when more than two priority queues and/or more hops have to be configured. 

Grigorjew et al.~\cite{grigorjew_ML_2020} highlight the fact that brute-forcing the optimal values does not scale for industrial use-cases. Therefore, they employ different machine-learning techniques and show that they can determine fitting delay bounds in the network. However, they assume that the flows' characteristics at each hop are known in advance, which removes the flexibility of online admission control schemes. Besides, in their evaluation, each hop in the network is configured identically, which may lead to potentially sub-optimal results.

In real applications, a set of static flows is often given, but operators might also wish for some flexibility to adapt the network during runtime. To this end, we propose our combined approach, where an offline optimization derives delay bounds for the online admission control.
For our proof-of-concept, we use~\cite{maile_journal_2022} to demonstrate the applicability of our solution to be used with online admission control. However, it should also be mentioned that this concept can be applied to any admission control scheme which requires delay bounds as input, such as~\cite{Guck,maile_journal_2022,grigorjew_decentralSP_2020,maile_decentral_2023}.
We use \ac{NC} - as it is a well-established delay analysis framework - to validate the delay bounds when flows are added to the network. NC models complex communication systems to derive performance guarantees, e.g., maximum per hop and end-to-end delays. To achieve this, worst-case maximum arrival and minimum departures of traffic are modeled using cumulative functions and combined using expressions from the min-plus algebra~\cite{boudec_network_2012}. For an introduction to \ac{NC}, see~\cite{boudec_network_2012}. We provide an overview of NC models for \ac{TSN} in~\cite{maile_network_2020}.

\begin{table}[t]
\centering
\begin{tabular}{|l|c|c|c|}
\hline
              & Sending Interval &  Max. Frame Size & Max. Latency\\
\hline
Profile 1 & \SI{250}{\micro\second} & 64B & \SI{250}{\micro\second} \\
Profile 2 & \SI{500}{\micro\second} & 128B & \SI{500}{\micro\second} \\
Profile 3 & \SI{1000}{\micro\second} & 256B & \SI{1000}{\micro\second} \\
Profile 4 & \SI{2000}{\micro\second} & 512B & \SI{2000}{\micro\second} \\
Profile 5 & \SI{4000}{\micro\second} & 1024B & \SI{4000}{\micro\second} \\
\hline
\end{tabular}
\vspace{2 mm}
\caption{Traffic profiles for industrial scenarios~\cite{grigorjew_ML_2020}}
\label{table:profiles}
\end{table}

\section{Network Configuration Framework for TSN}
\label{sec:TSN}
We present a framework for the configuration of TSN networks, which provides safe delay guarantees in networks which combine static and dynamic traffic flows. We refer to \textit{offline configuration} as the configuration of bridge parameters, e.g., the number of priority queues and the worst-case per-hop delay bounds. With \textit{online configuration} (or admission control), we refer to the registration and de-registration of flows while the network is running, after checking the flow's path for sufficient resources. Our goal is to offer delay-guaranteeing networks with only minimal required manual input. Therefore, we defined five traffic profiles for flows in Table~\ref{table:profiles}~\cite{grigorjew_ML_2020}. The profiles are derived from PROFINET use-cases for industrial sensor-controller networks. We use the term stream and flow interchangeably. 

\textit{Input:} The only input that we require is 1) the network topology, 2) the maximum number of priority queues, and 3) the percentage of bandwidth that shall remain free for future flow reservations. The latter is provided per link and per traffic profile. In addition, our approach can be optionally supplied with a set of static flows. 

\begin{figure}[t]
		\centering\includegraphics[width=0.55\linewidth]{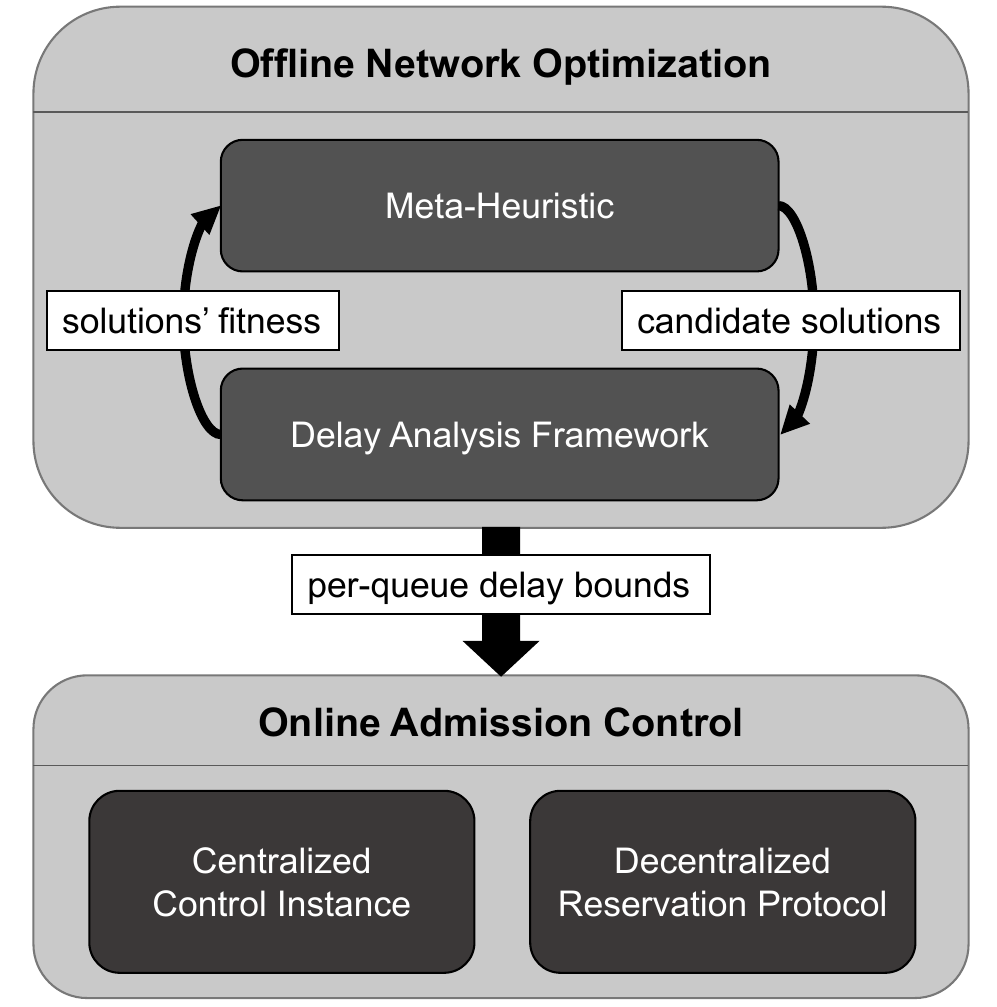}
		\caption{Framework overview. We use network optimization to derive delay bounds which improve the performance of online admission control.}\label{fig:overview}
\end{figure}

\textit{Offline Network Optimization:} We then run a meta-heuristic optimization which determines the number of required priority queues per hop, as well as the delay bounds for each priority queue in the network. This is illustrated in Fig.~\ref{fig:overview}.
Our heuristic generates potential candidate solutions, which define the delay bounds per priority queue. Each candidate solution is then evaluated according to its potential to provide real-time latency guarantees for the defined traffic profiles. This is expressed as a fitness value, which is defined by the objective function in Section~\ref{sec:objective}. The latency guarantees are validated using a delay analysis framework. For our tool, we implemented the delay analysis framework using NC, as NC allows for the efficient analysis of large and complex networks. The details of our delay analysis using NC can be found in \cite{maile_journal_2022}. The heuristic then iteratively aims at maximizing the fitness value returned by the objective function.
We compare two meta-heuristic algorithms, \ac{PSO}~\cite{pso} and \ac{GA}~\cite{ga_book}, as they have been widely used in various fields. 

\textit{Online Admission Control:} 
The best candidate solution then defines the delay bounds for each bridge, which are used to configure the network. These bounds remain static, while the actual flow reservations can change adaptively to the network setup, with no re-configuration of the network required. To add new flows to or remove existing flows from the network, existing online admission control algorithms are used, as proposed in~\cite{Guck,maile_journal_2022,grigorjew_decentralSP_2020,maile_decentral_2023}.

Our framework is suitable for centralized and decentralized network architectures, utilizing protocols specified in the respective standards. In a centralized network, the configuration is done in a so-called Central Network Controller (CNC), as defined by the IEEE 802.1Qcc-2018 standard~\cite{Qcc}. The CNC is a separate instance responsible for the reservation and configuration of all flows and nodes in the network. It is aware of the complete network topology, similar to a Software-Defined Networking controller. To communicate with the TSN network devices, network management protocols
are used such as the Network Configuration Protocol (NETCONF) defined in RFC 6241\footnote{\url{https://datatracker.ietf.org/doc/html/rfc6241}}. For decentralized networks, the Stream Reservation Protocol (SRP) has been proposed for online admission control in CBS networks. However, as we proved in~\cite{maile_decentral_2023}, the delay values proposed for SRP do not cover the worst-case. In 2018, the IEEE TSN group started working on the successor of SRP, the \ac{RAP}~\cite{Qdd}. Currently, RAP is still undergoing active development and is available in draft version 0.6. RAP considers introducing a pre-configuration phase of the network prior to online admission. This aligns perfectly with the requirements of our framework. Although the existing standards define the necessary protocols, they currently lack a solution for implementing resource allocation with guaranteed delays. Our framework fills this gap, and we hope that our research will help future standardization processes.

\begin{figure}[t]
	\centering
	\includegraphics[width=0.55\linewidth]{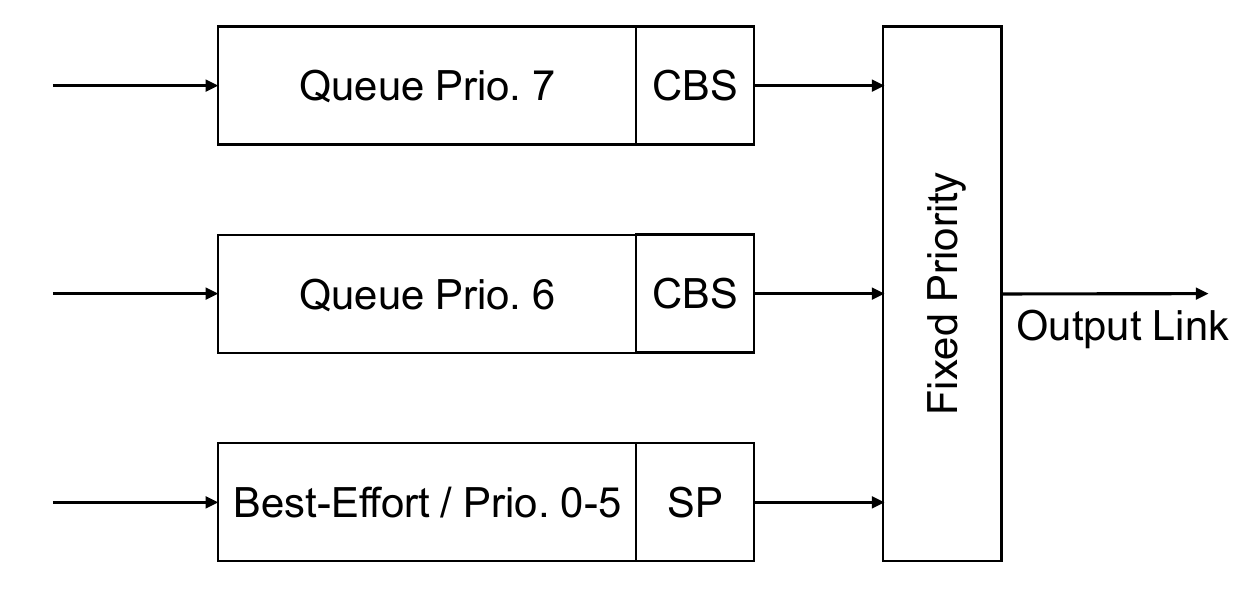}
	\caption{Example output port with CBS queues.}
	\label{fig:port}
\end{figure}

\section{Meta-Heuristic Optimization}
\label{sec:solution}

To derive the per-hop delay bounds for the network, we compare two meta-heuristic optimization approaches, \ac{PSO} and \ac{GA}. Initially, both algorithms start with a population of random solutions, which we call candidate solutions. The heuristics then successively update their solutions to improve their fitness. Eventually, the heuristics converge to a (near-optimal) solution.

TSN networks allow for up to eight queues in each output port of a bridge. Each queue is configured with a transmission selection algorithm. If multiple queues have eligible packets at the same time, the transmission is defined by the queues' priority. Fig.~\ref{fig:port} illustrates this for an example output port with two CBS queues. We define a subset of these queues as \textit{priority queues}, meaning that our framework optimizes them for time-critical traffic with deadline guarantees using CBS. All other queues can then be used by non-time-critical traffic without admission control and are not included in our optimization. 

\subsection{Design Decisions} 
\label{sec:design}
Our framework includes the following considerations, to ensure that a user only has to provide a minimum of input.

\textit{Routing:} We use a k-shortest delay-constrained routing algorithm~\cite{yin_k_shortest_routing}, with the links' bandwidth utilization as weight-function. We chose to weight the bandwidth utilization as this balances the network load, and thus, prevents bottlenecks which would decrease the possibility of online flow reservations.

\textit{Flow to Priority Assignment:} We do not require any kind of flow-to-priority mapping. Instead, our frameworks assign the priority to each flow automatically, by using the k-shortest paths and checking the priorities on each path with an end-to-end delay that is closest to a flow’s deadline successively. If no path allows for a reservation, the fitness of the solution decreases. To increase flexibility, we allow for individual per-hop priorities for each flow, instead of mapping a flow to a single priority on its whole path.

\textit{Number of Priority Queues:} The required number of priorities in a network is an optimization itself. We only require the definition of a maximum number of priority queues instead. Our heuristic will try to find a schedule with no high-priority queue unused, but potentially without using low priorities. Then, unused queues can be left out or used for other purposes, such as non-time-critical traffic.%

\textbf{Discussion:} 
The above considerations do not have any claim on optimality. E.g., instead of balancing the network, we could add the routing decision as part of the solution, thereby offering it as a variable for the optimization. All of these problems have been covered in heuristics themselves (e.g.,~\cite{li_sdn-based_2019,chuang_online_2020,frances_using_2006,soni_efficient_2020}), and including them increases the solution space and the runtime significantly. Our results show that with the above decision, already highly practical networks can be built. 

Note that we also allow for more detailed modeling, e.g., on the path of flows that arrive during online reservation. The more information is available, the better our approach can configure the network. However, to be appealing for real-life scenarios, we wanted to reduce the minimum required input as much as possible.

\subsection{Multi-Objective Function}
\label{sec:objective}
Each candidate solution is evaluated based on its capability of reserving all static flows and on the flexibility it offers for future reservations. The heuristics aim at maximizing the following fitness function, defined for each candidate solution $s$ as
\begin{equation}\label{eq:reward}
f(s) = \omega_1 \cdot f_R(s) + \omega_2 \cdot f_A(s) + \omega_3 \cdot f_D(s),
\end{equation}
where $\omega_i (i=1,2,3)$ define the weight for each of the objectives. We normalized the fitness value, so $f$ and $f_x (x=\{R,A,D\})$ are between 0 and 1. 

Thereby, $f_R(s)=1$, if all flows from the (optional) set of static offline flows can be reserved with the solution $s$. $f_R(s)$ can be derived by checking for each static flow whether a path can be found which will meet their end-to-end delay requirement. The ratio of successful reservations then defines $f_R(s)$. Even with a suboptimal solution, the flows' deadlines are still safe. We only reduce the maximum number of reservable flows in the network.

$f_A(s)=1$, if it is possible to reserve the required percentage of bandwidth for the online flows. We determine $f_A$ by artificially increasing the arrival function at each link for which the user assigned additional bandwidth for a traffic profile. This is done by dividing the end-to-end delay requirement of each profile into smaller per-hop delays, using the maximum path length of the network. We then check up to which arrival rate we can still guarantee the specified per-hop delay. As we are not aware of the path of the future flows and, thus, do not know the links from which they will arrive at each hop, we use the maximum rate that is possible for each link. $f_A$ then reflects the maximum ratio of available bandwidth when compared to the user's input.

\begin{figure}[t]
	\centering
	\includegraphics[width=0.45\linewidth]{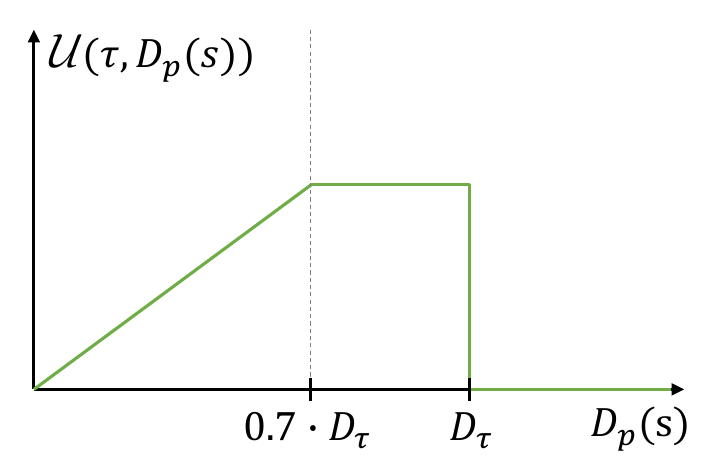}
	\caption{Utility function for flow $\tau$ with delay requirement $D_\tau$ and path delay $D_p$}
	\label{fig:utility}
\end{figure}

Finally, $f_D(s)=1$, if the end-to-end delay for the set of offline flows matches their deadline (within a $30\%$ flexibility interval). For each flow $\tau$, we define a utility function $\mathcal{U}(\tau, D_p(s))$~\cite{gavrilut_traffic-type_2020} to ensure that its end-to-end delay requirement, $D_\tau$, matches the path delay $D_p(s)$ for the solution $s$. This is illustrated in Fig.~\ref{fig:utility}. $f_D(s)$ is then defined as
\begin{equation}
f_D(s) = \frac{\sum_{\tau \in \mathcal{F}}\mathcal{U}\big(\tau,D_p(s)\big)}{|\mathcal{F}|},
\end{equation}
where $\mathcal{F}$ is the set of static flows. The reason for $f_D$ is that, if flows are scheduled faster than required, more bandwidth than necessary is reserved, which reduces the performance of best-effort traffic.
We define a candidate solution to be invalid, if some bridges in the network only utilize their low-priority queues, whereas the high-priority queues cannot be used for time-sensitive traffic. In these solutions, the high-priority queues would be configured unnecessarily. Therefore, these invalid solutions will be rewarded with a fitness value of 0.

We prioritize the individual objectives in our fitness function as follows: We assume that the successful reservation of the static offline flows is of the most importance. Afterwards, we optimize the bandwidth for future reservations of time-sensitive flows. Only if the above two goals are not hindered, the solution might be improved for best-effort and lower priority traffic. As a result, we chose the weight functions to reflect this prioritization as defined in Table~\ref{table:parameters}.

\subsection{Individual Per-Hop Delays}
\label{sec:heuristic}
The offline optimizations introduced in Section~\ref{sec:problem}, \cite{grigorjew_ML_2020,gavrilut_traffic-type_2020}, configured each hop in the network identically. We evaluate whether individual configurations for each network device can improve the networks' performance. As individual per-hop delays significantly increase the solution space, the heuristics potentially have a higher chance to remain in local optima, instead of finding the global best solution. Therefore, we propose a new approach, where we initialize the solution for individual delays with the results from the uniform network configuration. With this approach, the individual heuristics is guaranteed to perform equally or better than the uniform approach. We will show that this initialization can significantly improve the performance of the algorithms. 

\begin{table}[t]
\centering
\begin{tabular}{|l|l|}
\hline
              & \textbf{Values} \\
\hline
\textbf{PSO parameters} &   \\
$w$, $c_1$, $c_2$            & 0.5, 2, 2.4  \\
\hline
\textbf{GA parameters}           &   \\
crossover operation     & blend crossover, $P_{c}=0.45, \alpha=0.15$  \\
mutation operation     &  NSGA-II, $P_{m}=0.45, \eta=70, P_{ind}=0.3$  \\
selection algorithm     & NSGA-II, $P_s=0.5$  \\
\hline
\textbf{Shared parameters}           &   \\
$\omega_1$, $\omega_2$, $\omega_3$      & 0.9, 0.09, 0.01  \\  
population size, convergence      & Uniform config.: 100, 15; Individual config.: 200, 20\\
\hline
\end{tabular}
\vspace{2 mm}
\caption{Parameter selection for meta-heuristics}
\label{table:parameters}
\end{table}

\subsection{Parameter Choice}
\label{sec:parameters}

We have implemented both, \ac{PSO} and \ac{GA}, using Python3. For the \ac{GA} implementation, we used the evolutionary computation framework DEAP~\cite{DEAP_JMLR2012}. To determine generic and efficient parameters for both approaches, we used the hyperparameter optimization framework Optuna~\cite{optuna_2019}. To prevent overfitting, we created an evaluation set with highly variable network topologies, different numbers of priorities, and changing flow characteristics. The topologies are presented in Section~\ref{sec:evaluation}. 

\ac{GA}s are inspired by the principles of biological evolution and natural selection. The candidate solutions are updated using selection, crossover, and mutation operators. Selection involves choosing a percentage of $P_s$ individuals from the current population based on their fitness values. Crossover and mutation then adapt a certain percentage of individuals ($P_c$ and $P_m$, respectively) to explore the solution space. We evaluated all available GA mutation, crossover, and selection algorithms in DEAP, along with their parameter values. 

Similarly, \ac{PSO} iteratively updates the candidate solutions based on the solution's own experience and the collective knowledge of the population. The update uses an inertia $w$, a cognitive component $c_1$, and a social component $c_2$  as parameters. The inertia term allows particles to maintain their momentum, while the cognitive component focuses on the particle's personal best position, and the social component emphasizes the global best position.

The best set of parameters is provided in Table~\ref{table:parameters}. We also use them for our evaluation. 

\section{Evaluation}
\label{sec:evaluation}
We have defined three different topologies which are typical for sensor-controller networks in the industry. Controllers refer to Programmable Logical Controllers (PLCs) and represent end-stations which control industrial machines, e.g., in manufacturing processes.  We assume 1 Gbit/s links in all topologies and test cases. We used the following topologies, as they are typical for industrial use-cases: 1) a \textit{line} topology with four bridges, an end-station connected to each of them, and a PLC at the end of the line, 2) a \textit{star-of-stars} topology with one central bridge connected to four bridges, which are again connected to four end-stations, 3) a ring topology with five bridges and each bridge connected to two end-stations. For topologies 2) and 3), we randomly assign one end-device to be the PLC. In all scenarios, flows are addressing the PLC in the network to create a bottleneck for our evaluation.
We defined a discrete search space $S$ with steps of \SI{10}{\micro\second} between $0$ and \SI{4}{\milli\second}, for all algorithms, as \SI{4}{\milli\second} is the maximum possible deadline for our flows. 
\subsection{Benchmark Algorithms}
We have implemented two approaches as benchmark algorithms for our evaluation. Thereby, we want to evaluate both, the increase in optimality when compared to an intuitive configuration and the gain of performance when compared to an exhaustive search. We evaluate our solution in both categories by using the following two approaches. For comparison, all code is executed without parallelization on an Intel Xeon Silver 4215R processor with 3.20 GHz.

\subsubsection{Exhaustive Search (ES):}
The exhaustive search will iteratively evaluate all possible solutions, and thus, can determine the optimum within the search space. With this approach, we can evaluate how close our results are to the optimal value. However, the exhaustive search suffers from a high runtime. Thus, it cannot cover individual per-hop delays, but will only investigate settings with identical delays for all hops. We will show in our evaluation that, due to this limitation, our solution is able to achieve even better configurations than the exhaustive search.

\subsubsection{Intuitive Approach (IA):}
Our intuitive approach reflects the configuration of a user. It will serve as a benchmark to evaluate whether our heuristics are actually needed to configure high-performance networks. Simply spoken, the intuitive approach will uniformly distribute the end-to-end delays of each flow over its number of hops on the path. The resulting values represent the per-hop delays which each flow requires to meet its deadline requirement. We then configure the network by deriving the quantiles from this set in a way that each flow can be covered by one of the queues. E.g., for four queues, we use the minimum per-hop value, plus the 25\%, 50\%, and 75\% quantile of the resulting delays to configure the four queues. When compared to our solution, this approach has only minimum runtime, but it does not consider the effect that traffic load has on the queuing delays in the network. As a result, our solution is evaluated against the intuitive approach, i.e., to check for improved fitness values alias more successful flow reservations.

\subsection{Convergence}
\begin{figure}[t]
		\centering\includegraphics[width=0.52\linewidth]{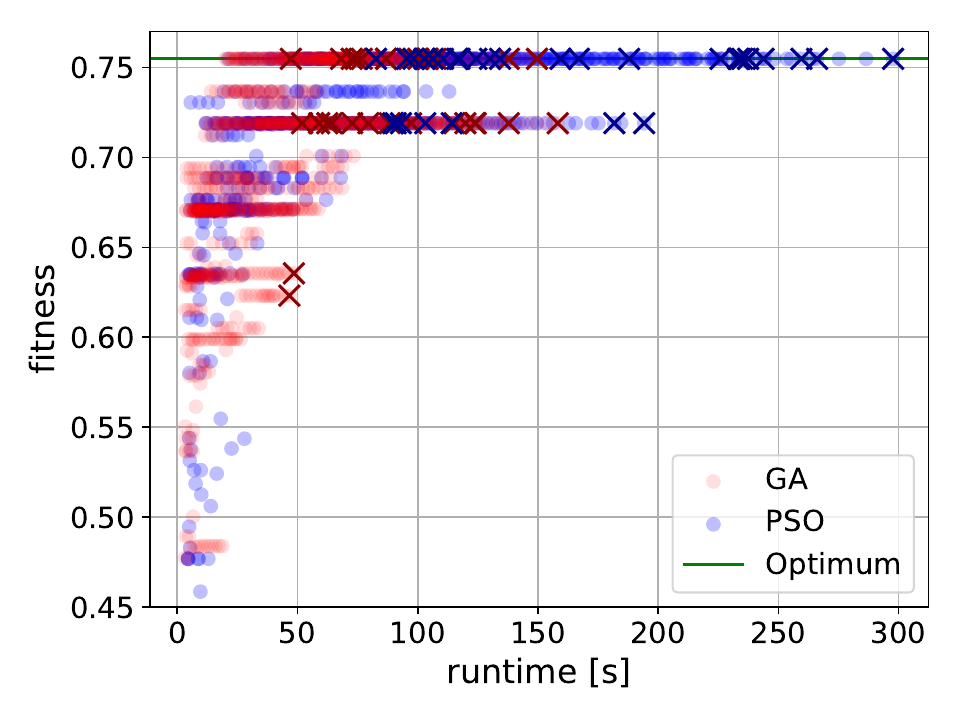}
		\caption{Results of GA and PSO after each iteration, with results at termination marked as X. The optimum is derived using ES.}\label{fig:convergence}
\end{figure}

Fig.~\ref{fig:convergence} illustrates the behavior of the GA and PSO algorithm when compared to the ES in the star-of-stars network, with 150 flows randomly assigned to traffic profiles and end-stations. We assumed a maximum of two CBS queues and a uniform configuration for all network devices. With these settings, the ES can obtain the optimum in \SI{119}{\minute}, and we can see how the heuristics perform in comparison. Fig.~\ref{fig:convergence} shows the result after each iteration for 30 independent replications, where we consider the algorithms as converged if their results remain constant for 10 iterations. The final result after the algorithms have converged is marked with an X. As we can see, both heuristics achieve good results, but GA is faster in reaching high fitness values than PSO. GA performs faster as it only updates solutions with a specific probability ($P_c$ and $P_m$ of Table~\ref{table:parameters}). However, GA is twice as likely to converge in a local rather than global optima when compared to PSO.

\subsection{Comparison}

To compare our heuristics, we conduct experiments by uniformly choosing one of the three topologies. We randomly generated 200 flows uniformly distributed with the profiles defined in Table~\ref{table:profiles}, originating from random sources and heading to the PLC of the network. We repeat the experiments 30 times for each algorithm. We first assume that all hops are configured with the same delay values.

\begin{figure*}[t]
	\centering
	\includegraphics[trim=10pt 40pt 10pt 30pt,clip,width=0.495\linewidth]{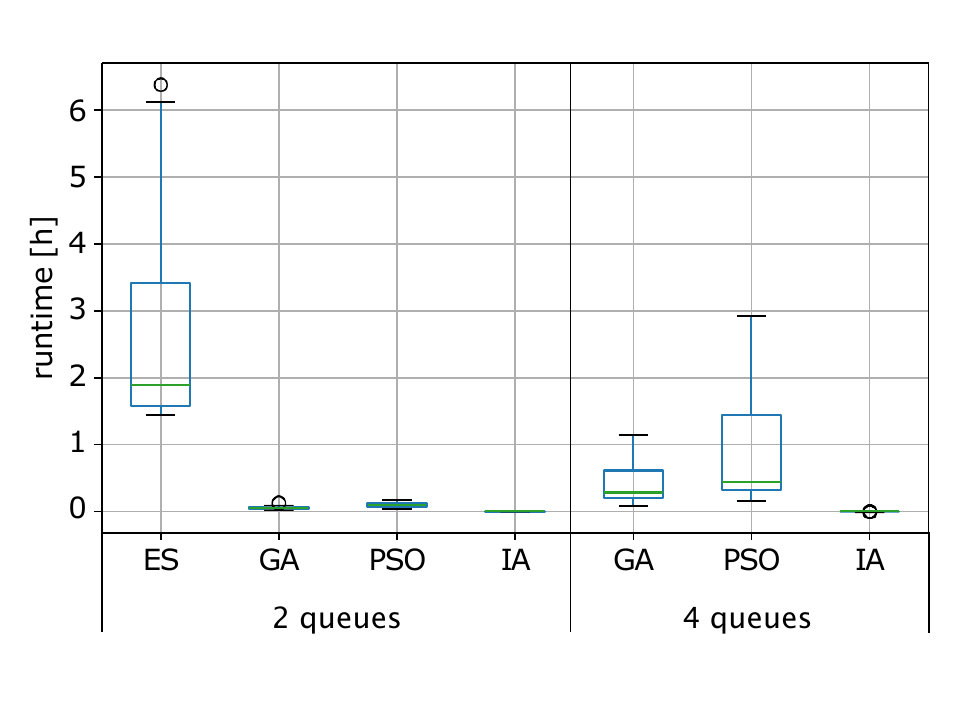}
 	\includegraphics[trim=10pt 40pt 10pt 30pt,clip,width=0.495\linewidth]{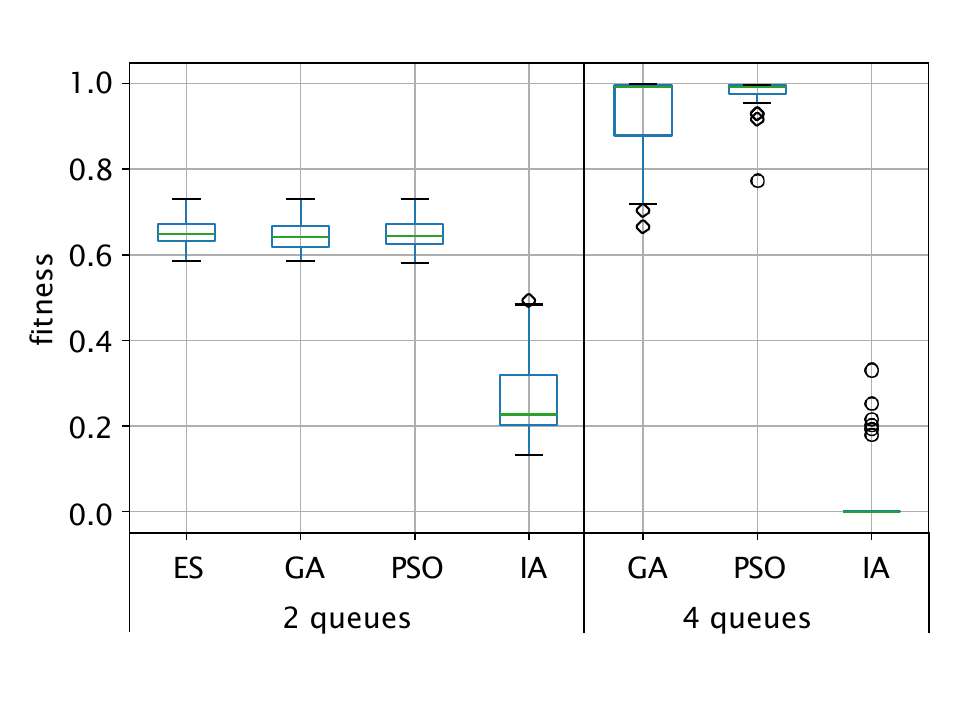}
	\caption{Runtime and fitness results in networks with uniform bridge configuration.}
	\label{fig:comparison}
\end{figure*}

Fig.~\ref{fig:comparison} shows the runtime and resulting fitness values for the experiments, respectively. With the given search space and two priorities, the ES has to evaluate $|S|^2\approx160\cdot10^3$ possible solutions. Increasing the number of priorities to four would result in $|S|^4\approx25\cdot10^9$ solutions, which cannot be accomplished by the ES due to its performance. 

As we can see in Fig.~\ref{fig:comparison}, both heuristics can achieve similar fitness values as the ES for two priorities. While the intuitive approach does allow for some reservations, it cannot compete with the results from our heuristics. For two priorities, the heuristics perform 2.4 times better on average. Additionally, the heuristics can evaluate more priorities than the ES, which results in more successful reservations. E.g., for four priority queues, GA performed 41.4\% and PSO 49.3\% better than the ES with two priorities. Again, the GA provides better performance in terms of runtime than the PSO but is more likely to converge in local optima. For the four priority settings, the intuitive approach frequently results in invalid configurations, where high-priority queues remained unused.

\begin{figure}[t]
		\centering\includegraphics[width=0.4\linewidth]{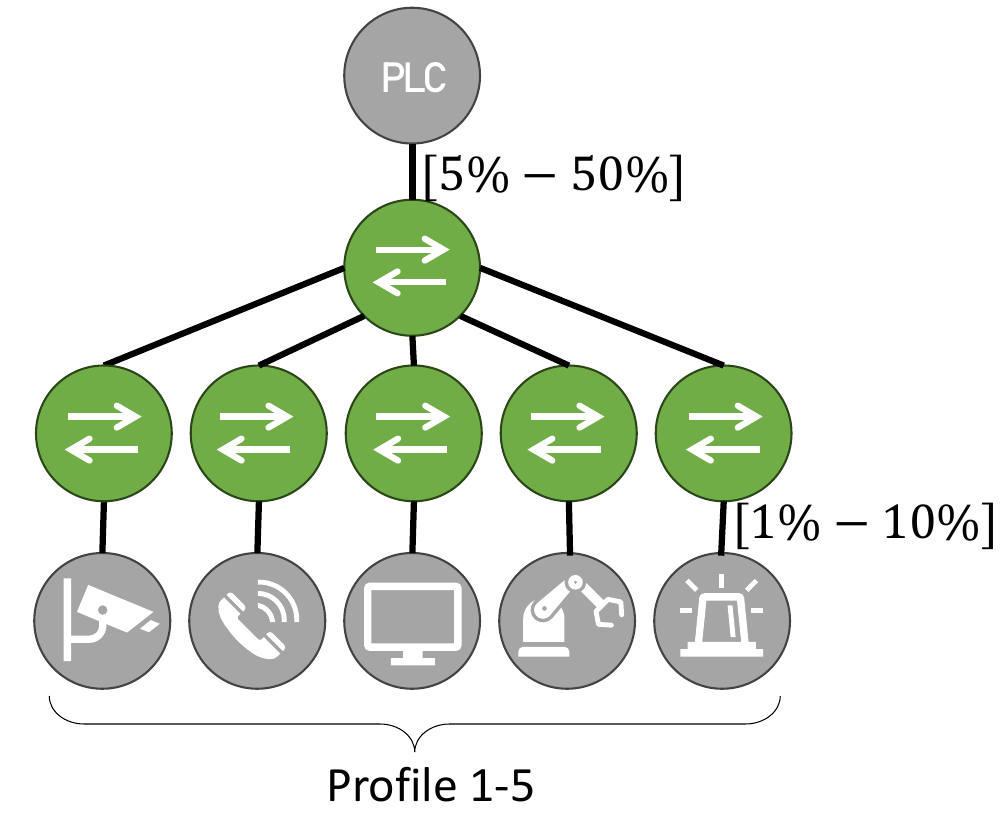}
		\caption{PROFINET network with given link utlizations.}\label{fig:profinet_network}
\end{figure}

\subsection{Individual per-hop Delays}
Individual per-hop delays allow for a better distribution of the network resources when the traffic load is not evenly distributed. For this, we defined the network shown in Fig.~\ref{fig:profinet_network}, based on the PROFINET design guidelines~\cite{profinet} with one priority per flow. Each line is connected to an end-station with an individual traffic profile. We randomly generated flows to simulate a link utilization between 5\% and 50\% on the link to the PLC.

\begin{figure*}[t]
	\centering
	\includegraphics[trim=10pt 70pt 10pt 30pt,clip,width=0.495\linewidth]{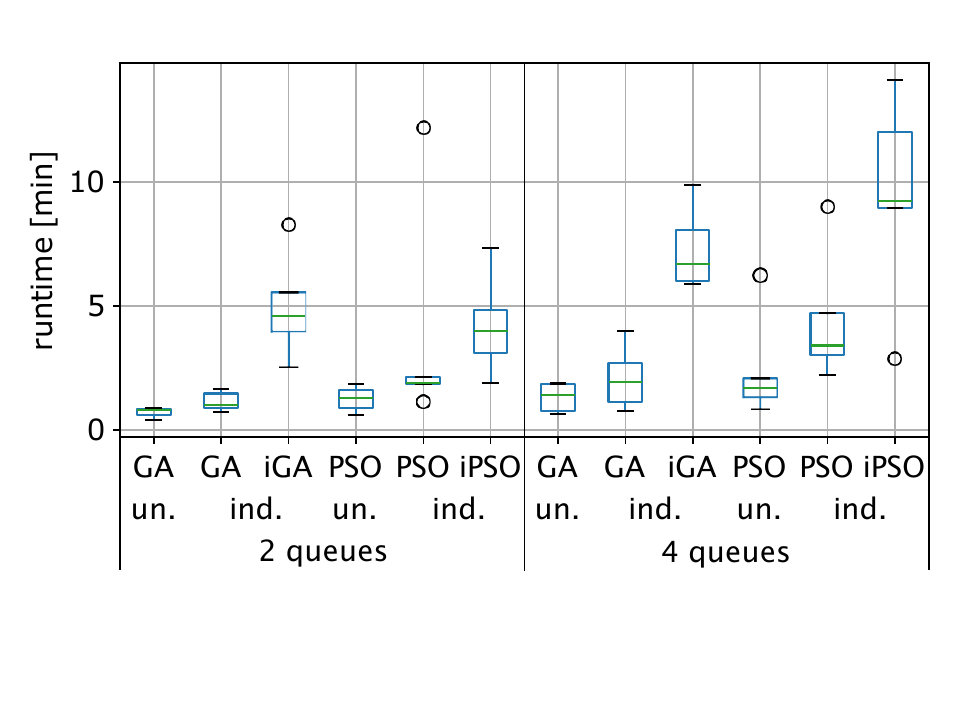}
 	\includegraphics[trim=10pt 70pt 10pt 30pt,clip,width=0.495\linewidth]{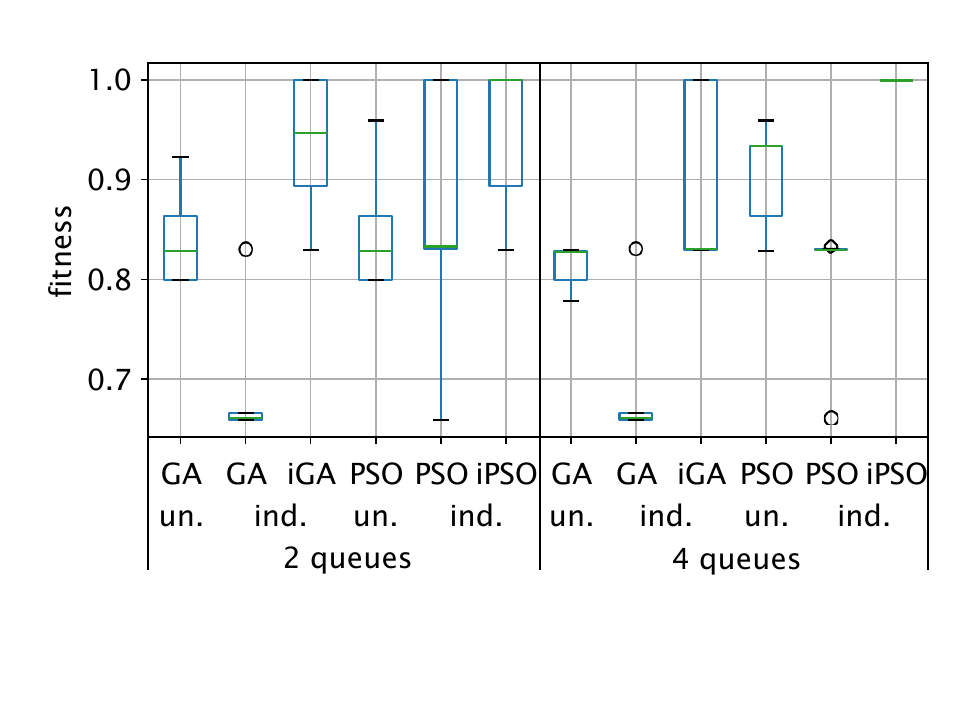}
	\caption{Runtime and fitness results in a network with individual per-hop configuration.}
	\label{fig:comparison_per_hop}
\end{figure*}

While neither the ES nor the intuitive approach allow for individual per-hop delays, both of our heuristics can evaluate such configurations. Fig.~\ref{fig:comparison_per_hop} shows the result for the GA and PSO algorithms. We compared the uniform network configuration ('un.') with the individual configuration ('ind.'). In addition, we show the results for our initialized solutions ('iGA' and 'iPSO'), which we initialize with the results from the uniform configuration to determine individual delay bounds. We can see that simply allowing for individual queue delays can reduce the performance of the heuristics when compared to the uniform configuration results. This is due to the vast increase of the solution space, which makes it more likely for the heuristics to converge in local optima. 
However, our initialized algorithms iGA and iPSO can improve the results from the uniform configuration by 10-11\% on average and, at the same time, are guaranteed to not provide lower results.

\subsection{Flows during Runtime}
\begin{figure}[t]
		\centering\includegraphics[width=0.6\linewidth]{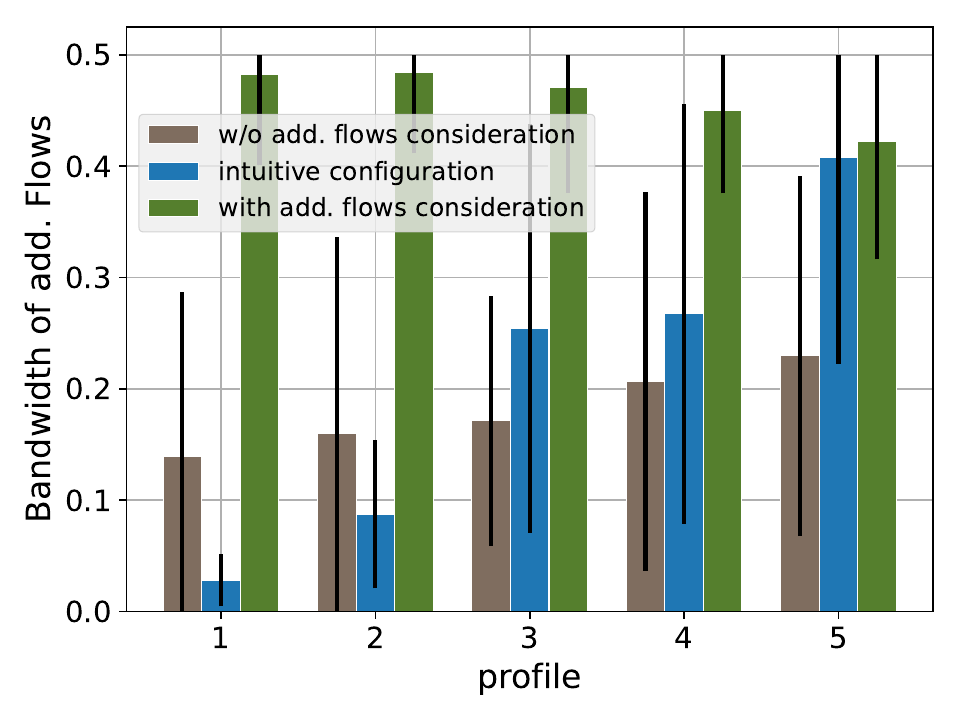}
	\caption{Successful flow reservations with online admission and one standard deviation. In separate networks, each profile should receive 50\% bandwidth.}\label{fig:additional_flows}
\end{figure}

Finally, to evaluate the benefit for future flow reservations, we again use the star-of-stars topology and define 10 offline flows for each of the profiles 1, 3, and 5. We run different scenarios, where in each scenario, we reserved 50\% of the bandwidth for one of the five traffic profiles for future flows. For each scenario, we repeated the experiment 10 times, resulting in 50 repetitions in total. Fig.~\ref{fig:additional_flows} compares the number of successful reservations during online admission, when the offline optimization considers the bandwidth for the future flows, in comparison to when they are not considered. Additionally, we also compared our heuristics to the intuitive configuration. For these scenarios, we improved the intuitive approach by artificially adding every possible future flow.
As we can see, all configurations have some potential to add additional flows during the runtime of the network. However, our combined solution which considers the required bandwidth for future flows can increase the number of successful reservations when compared to the intuitive approach and a configuration without future flow considerations significantly, illustrating our framework's flexibility to support dynamic traffic during runtime. Fig.~\ref{fig:additional_flows} also shows that flows with tight deadline guarantees (e.g., profile 1) profit most from our optimization, while higher latency requirements (e.g. profile 5) are easier to integrate even without specific considerations.

\section{Conclusion}
\label{sec:conclusion}
We have presented our framework for offline and online configuration of TSN networks with delay guarantees. Thereby, the offline configuration allows for efficient usage of the network resources, while the online configuration allows for the registration and deregistration of flows while the network is running. By considering future flows in the offline configuration phase, we ensure that the network can easily react to changing network conditions while still offering guarantees for the end-to-end delays of time-critical flows. Our framework requires only minimal user input, making it highly relevant for practical application and providing a first step towards the auto-configuration of TSN networks.

We have evaluated two meta-heuristics for our framework and showed that a particle swarm optimization provides better results when compared to a genetic algorithm, with slightly higher runtimes. We also implemented two benchmark algorithms, one exhaustive search to evaluate the optimum for simple configurations, and one intuitive approach to reflect the behavior of a user. We could show that our heuristics reached similar results as the exhaustive search in just a fraction of the time. 
Due to the high flexibility and the low runtime, our framework can provide even better results than the exhaustive search and significantly better results than an intuitive solution. We demonstrated that allowing individual delay bounds on each hop improves the network's performance. Finally, we also showed the effect on the success of new flow reservations while the network is running. Thereby, considering the required bandwidth of future flows during the offline configuration highly increases the chance for future flow reservations.

As future work, we want to extend the delay-guaranteeing admission approaches as proposed by~\cite{grigorjew_decentralSP_2020,Guck,maile_decentral_2023,maile_journal_2022} to more schedulers. Specifically, we would like to cover solutions for the Asynchronous Traffic Shaper and Cyclic Queueing and Forwarding networks, to ensure that our framework is applicable for a wide range of TSN scenarios.
\bibliographystyle{splncs04}
\bibliography{TSN_u_NC}

\begin{acronym}
\acro{WCRT}[WCRT]{worst-case response time}
\acro{SFD}[SFD]{start frame delimiter}
\acro{IFG}[IFG]{inter-frame gap}
\acro{IPG}[IPG]{inter-packet gap}
\acro{SP}[SP]{Strict Priority}
\acro{QoS}[QoS]{Quality of Service}
\acro{TSN}[TSN]{Time-Sensitive Networking}
\acro{NC}[NC]{Network Calculus}
\acro{CMI}[CMI]{Class Measurement Interval}
\acro{MFS}[MFS]{Maximum Frame Size}
\acro{MIF}[MIF]{Maximum Interval Frame}
\acro{AV}[AV]{Audio and Video}
\acro{AVB}[AVB]{Audio and Video Briding}
\acro{SRP}[SRP]{Stream Reservation Protocol}
\acro{SR}[SR]{Stream Reservation}
\acro{CBS}[CBS]{Credit-Based Shaper}
\acro{BE}[BE]{Best-Effort}
\acro{ES}[ES]{Exhaustive Search}
\acro{RAP}[RAP]{Resource Allocation Protocol}
\acro{FIFO}[FIFO]{First-In-First-Out}
\acro{FoI}[FoI]{Flow of Interest}
\acro{NRT}[NRT]{Non-Real-Time}
\acro{PLC}[PLC]{Programmable Logic Controller}
\acro{SSRP}[RRP]{Reliable Reservation Protocol}
\acro{RAP}[RAP]{Resource Allocation Protocol}
\acro{NMS}[NMS]{Network Management System}
\acro{TSpec}[TSpec]{Traffic Specification}
\acro{PSO}[PSO]{Particle Swarm Optimization}
\acro{GA}[GA]{Genetic Algorithm}
\end{acronym}

\end{document}